\documentclass[aps,prl,twocolumn,showpacs,nofootinbib,superscriptaddress]{revtex4} 

\usepackage{graphicx,longtable,epsfig} %
\usepackage{dcolumn}
\usepackage{mathptmx}
\usepackage{amsmath}
\usepackage{epstopdf}
\usepackage[pdftex]{hyperref}
\usepackage{sidecap}
\sidecaptionvpos{figure}{c}
\usepackage[sort&compress]{natbib}

\newcommand{\beqy}{\begin{eqnarray}}
\newcommand{\eeqy}{\end{eqnarray}}
\newcommand{\bmlet}{\begin{subequations}}
\newcommand{\emlet}{\end{subequations}}
\newcounter{saveeqn}

\def\gsimeq{\,\,\raise0.14em\hbox{$>$}\kern-0.76em\lower0.28em\hbox  
{$\sim$}\,\,}  
\def\lsimeq{\,\,\raise0.14em\hbox{$<$}\kern-0.76em\lower0.28em\hbox  
{$\sim$}\,\,}  

\begin{document}

\title{Large low-energy $M1$ strength for $^{56,57}$Fe within the nuclear shell model}

\author{B. Alex Brown}
\affiliation{National Superconducting Cyclotron Laboratory and Department of Physics and Astronomy, 
Michigan State University, East Lansing, Michigan 48824-1321, USA}
\email{brown@nscl.msu.edu}
\author{A.~C.~Larsen}
\affiliation{Department of Physics, University of Oslo, N-0316 Oslo, Norway}
\email{a.c.larsen@fys.uio.no}

\date{\today}

\begin{abstract}
A strong enhancement at low $\gamma$-ray energies has recently been discovered in the 
$\gamma$-ray strength function of $^{56,57}$Fe. In this work, we have for the first time
obtained theoretical $\gamma$ decay spectra for states up to $\approx 8$ MeV in excitation for $^{56,57}$Fe.
We find large $B(M1)$ values for low $\gamma$-ray energies that provide an 
explanation for the experimental observations. The role of mixed $E2$ transitions for the low-energy enhancement 
is addressed theoretically for the first time, and it is found
that they contribute a rather small fraction.
Our calculations clearly show that the high-$\ell$ ($=f$) diagonal terms are most important for the strong low-energy 
$M1$ transitions. As such types of $0\hbar\omega$ transitions are expected for all nuclei, our results indicate 
that a low-energy $M1$ enhancement should be present throughout the nuclear chart. This could have
far-reaching consequences for our understanding of the $M1$ strength function at high excitation energies, 
with profound implications for astrophysical reaction rates.
\end{abstract}


\maketitle


Gamma-absorption and decay properties of atomic nuclei are of crucial importance in fundamental and
applied nuclear-physics research. They give information on the nuclear structure
and are indispensable for cross-section calculations for a broad range of applications, such as 
next-generation nuclear reactors and for the description of the nucleosynthesis in explosive stellar environments.

For $\gamma$-absorption cross sections above the particle thresholds, data are fairly complete for nuclei close to the
valley of stability~\cite{dietrich1988}, although still very scarce for exotic nuclei 
(see e.g. Refs.~\cite{adrich2005,rossi2013}). The Giant Electric Dipole Resonance (GDR) is the dominant feature
and its $E1$ strength overshadows all other decay modes for $E_\gamma \approx 12 -17$ MeV. Below the neutron threshold, 
the $\gamma$-ray strength function ($\gamma$SF), i.e. the average, reduced $\gamma$-decay probability, is not as well known
as the photoneutron cross sections, although more and more pieces to the full picture are emerging~\cite{ocl-website}.

Over the past 10 years, measurements on the $\gamma$SF of many 
\textit{fp}-shell~\cite{voinov2004,larsen2006,larsen2007,voinov2010,burger2012,larsen2013} 
and $A\sim90-100$ nuclei~\cite{guttormsen2005,wiedeking2012}
have revealed a surprising feature: the probability of $\gamma$ decay increases as the $\gamma$-ray energy 
decreases. Such a behavior is the complete opposite of what was expected from traditional $E1$ models, both 
semi-phenomenological approaches (e.g. Ref.~\cite{kopecky1990}) and  more microscopic ones
(e.g. Ref.~\cite{goriely2002}).
However, recent theoretical work on Mo isotopes show that a low-energy increase in the $\gamma$SF could be due to 
thermal single-quasiparticle transitions into the continuum, giving rise to enhanced $E1$ strength for low
$\gamma$-ray energies~\cite{litvinova2013}. On the other hand, shell-model calculations on $^{94,95}$Mo and $^{90}$Zr give
high $B(M1)$ values for low $\gamma$-rays caused by a spin re-coupling of high-$j$ proton and 
neutron orbits~\cite{schwengner2013}. 

The low-energy enhancement is very intriguing, as it may represent a completely new decay mode and reveal so-far unknown
nuclear-structure effects; as such, it
is being subject to intense research. Moreover, it may have far-reaching consequences for the rapid neutron-capture process,
the astrophysical nucleosynthesis responsible for creating $\approx$ 50\% of the nuclides in the solar 
system~\cite{burbidge1957,cameron1957};
the presence of an enhanced decay-probability for low-energy $\gamma$-rays may increase the ($n,\gamma$) reaction rates 
$1-2$ orders of magnitude~\cite{larsen_goriely2010}. As clearly expressed in Refs.~\cite{arnould2007,surman2014}, 
astrophysical ($n,\gamma$) rates are vital in sophisticated r-process models. 

In this Letter, we present the first large-basis shell-model calculations for the 
$\gamma$-decay spectra of levels up to excitation energies of
$\approx 8$ MeV in $^{56}$Fe and $^{57}$Fe. The calculations reveal a 
strong $M1$ component in the $\gamma$SF for low-energy $\gamma$-rays. 
The shape of the calculated $M1$ $\gamma$SF is in excellent agreement 
with the data of Refs.~\cite{voinov2004,larsen2013}. Moreover, we investigate the role of $E2$ $\gamma$-rays, as it was
found in Ref.~\cite{larsen2013} that a small contribution ($\approx 10$\%) of stretched $E2$ transitions could possibly
be present. Also, the mechanism behind the enhancement will be explained.

We used the GPFX1A Hamiltonian \cite{gx1a1,gx1a2} for the $  pf  $ shell.
Excitation energies obtained with this Hamiltonian in the region of
$^{56}$Fe are in excellent agreement with experimental
energies up to about 8 MeV when the $  J  $ value is known experimentally \cite{gx1a1}.
The model space for $^{56}$Fe was $  (0f_{7/2})^{6-t}(0f_{5/2},1p_{3/2},1p_{3/2})^{t}  $
for protons and $  (0f_{7/2})^{8-t}(0f_{5/2},1p_{3/2},1p_{3/2})^{t+n}  $ for neutrons,
where $  n=2  $ and $  t  =0$, 1 and 2. The lowest lying states are dominated by $  t=0  $, but
the core-excitations with higher $  t  $ values are required for states
up to about 8 MeV. With this model space there are a total of 6,046,562 states.
With the code NuShellX \cite{nushellx} the Lanczos method was used to obtain the
eigen-energies and eigen-vectors
for the lowest 50 states of each $  J  $ value with an accuracy of about 1 keV.
This provides a complete set of states within the model space up to about 7.5 MeV.
There are 255 positive-parity states up to 7.5 MeV for $^{56}$Fe.
This model space does not include the negative-parity
states that start experimentally with the 3$^{-}$ level at 4.37 MeV.

We calculated the complete set of $M1$ and $E2$ matrix elements
for the 50 positive-parity states of each spin, a total of about 10$^{5}$ matrix elements.
These were used to calculate lifetimes, branching ratios and mixing
ratios for the $\gamma$ decay. For $E2$ we used the standard effective
charges of $  e_{p}=1.5  $ and $  e_{n}=0.5  $. 
For the $M1$ transitions, we used the effective $M1$ operator of Ref.~\cite{effM1}.
The matrix elements for the $E2$
were obtained with harmonic-oscillator radial wavefunctions.

For $^{57}$Ni (with $  n=3  $) it was only possible to include $  t  =0$ and 1.
There are 233,793 negative-parity states in this model space for $^{57}$Fe.
This truncation for $^{56}$Fe reduces the number of states up to 7.5 MeV
by about 30\%.
The considered spin range was $J=0-10$ and $J=1/2-21/2$ for $^{56,57}$Fe,
respectively. 
 \begin{figure}[t]
 \begin{center}
 \includegraphics[clip,width=1.\columnwidth]{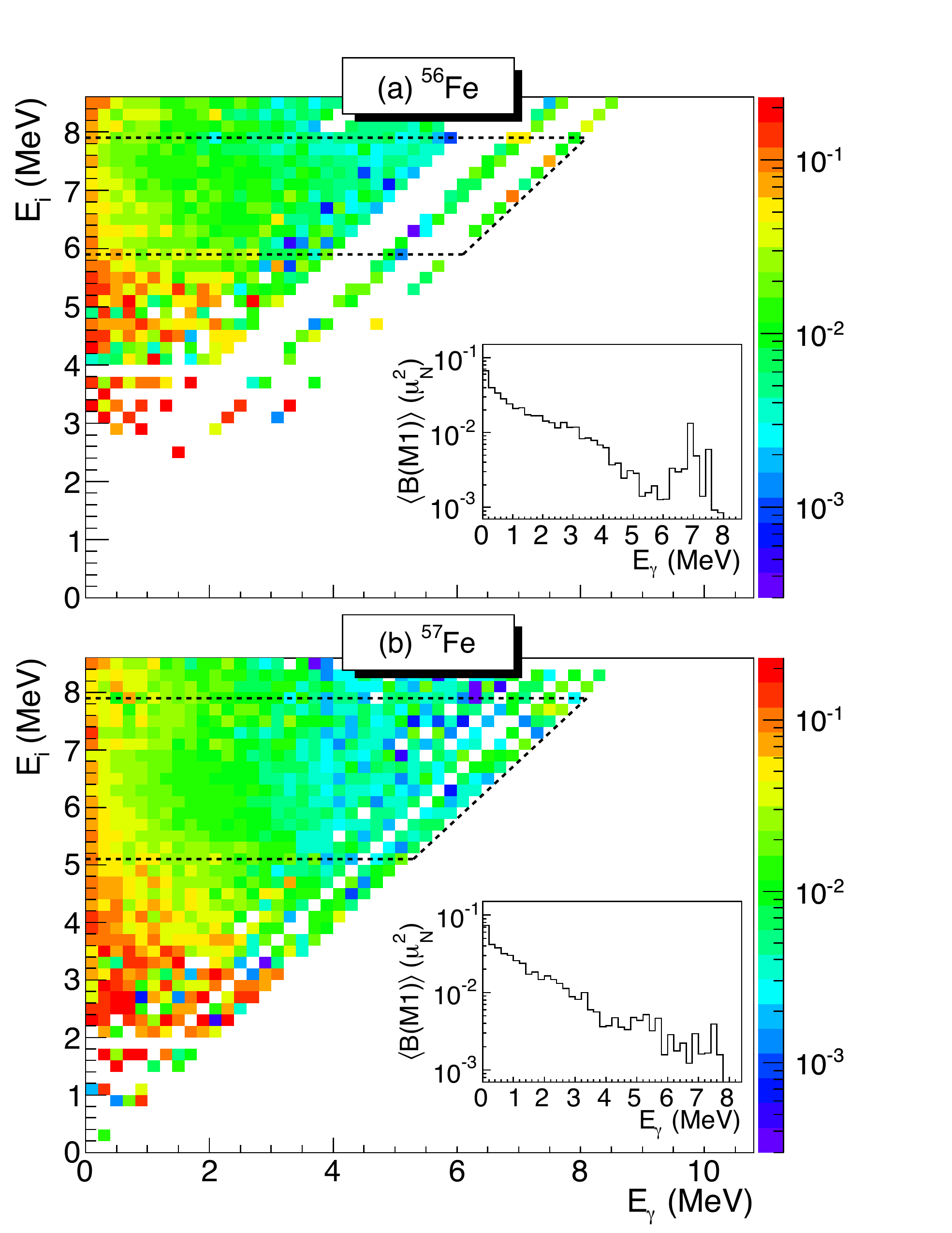}
 \caption {(Color online) Caluclated distributions of average
 $B(M1)$ values as a function of initial excitation energy $E_i$ and $\gamma$-ray energy $E_\gamma$ 
 for (a) $^{56}$Fe and (b) $^{57}$Fe. The insert shows the projection of the $\left<B(M1)\right>$
 values for the $E_i$ gates indicated in the figure. Note the log scale on the $\left<B(M1)\right>$ axis.}
 \label{fig:BM1_matrices}
 \end{center}
 \end{figure}

For each level, detailed decay information is available, such as the branching ratios, the magnetic dipole
and electric quadrupole transition strengths $B(M1)$ and $B(E2)$ for each individual transition, as well as the mixing ratio
$\delta$ defined as $\delta^2 = \lambda_{E2}/\lambda_{M1}$,
where $\lambda_{E2}$ and $\lambda_{M1}$ are the $E2$ and $M1$ transition rates. 
The calculated transitions were sorted into matrices with 200-keV wide energy bins, 
both for the initial excitation energy and the transition energy, incrementing the $B(M1)$ transition strengths. 
Moreover, we have calculated the
average $B(M1)$ $\gamma$-decay transition strengths for each ($E_\gamma,E_i$) pixel simply by dividing each pixel 
with the number of $M1$ transitions in that pixel in the same way as in Ref.~\cite{schwengner2013}. 
By sorting the information in this way, we obtain ($E_\gamma,E_i$) matrices that correspond to the experimental situation, 
such as the data of $^{56}$Fe, Fig.~3 in Ref.~\cite{larsen2013}. 
The ($E_\gamma,E_i,\left<B(M1)\right>$) matrices from the shell-model calculations are shown for $^{56,57}$Fe
in Fig.~\ref{fig:BM1_matrices} a and b, respectively. Note that the $B(M1)$ values are from both pure and 
$E2$-mixed $M1$ transitions. 

 \begin{figure}[t]
 \begin{center}
 \includegraphics[clip,width=0.95\columnwidth]{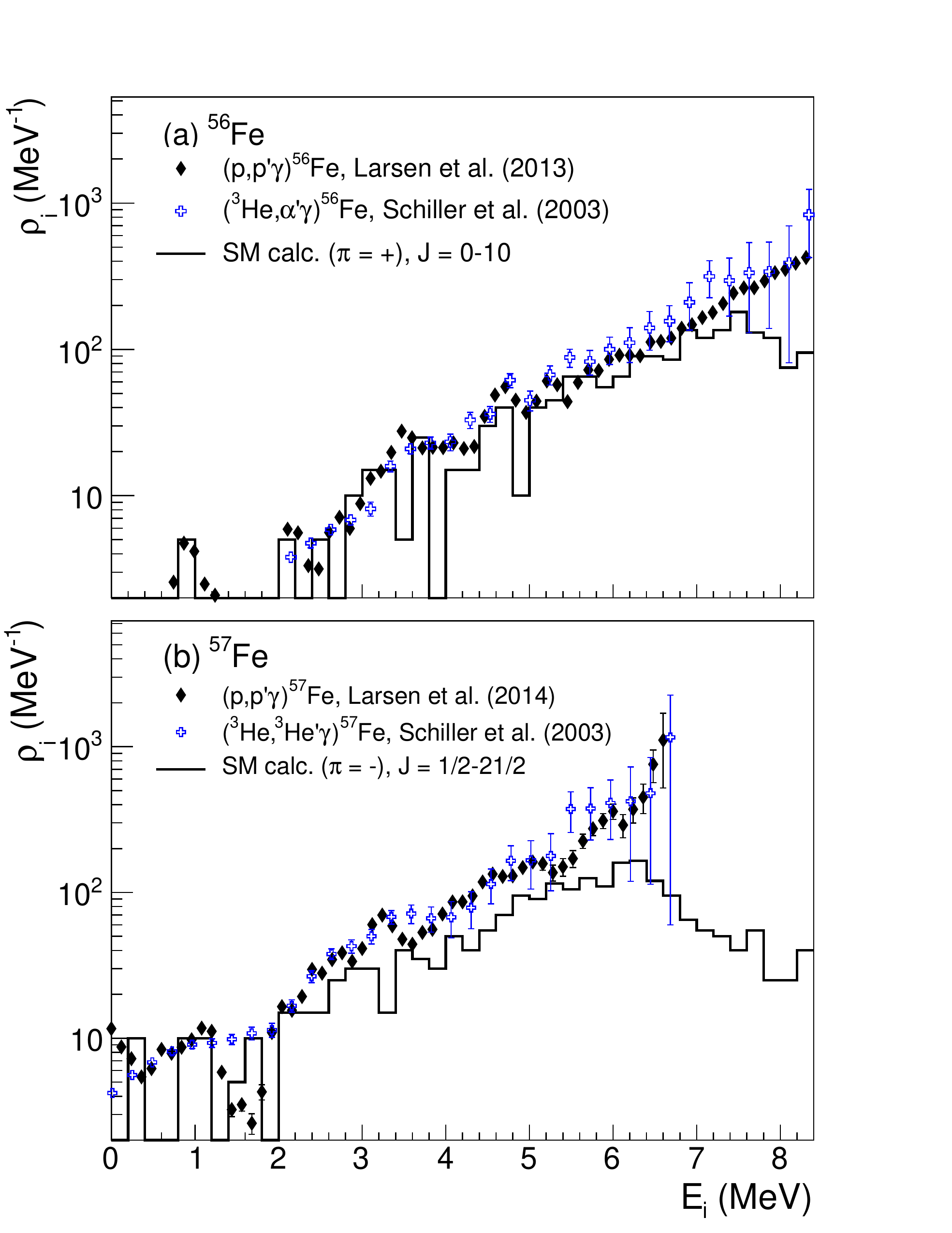}
 \caption {(Color online) Shell-model level densities (excited levels) for (a) $^{56}$Fe and (b) $^{57}$Fe, compared to 
 experimental data from Refs.~\cite{schiller2003,larsen2013,larsen2014}.} 
 \label{fig:nld}
 \end{center}
 \end{figure}

The obtained shell-model level densities are shown in Fig.~\ref{fig:nld} and compared to experimental
data (both parities). The theoretical level density for  $^{56}$Fe is a little lower than experiment due to 
the presence of negative parity states.  The theoretical level density for  $^{57}$Fe is lower than experiment also due to 
the truncation. To examine the effect of truncation we have also done $^{56}$Fe with the more
restrictive $t \leq 1$ truncation. Even though the level density becomes factor of two lower,
the $\gamma$SF obtained for $^{56}$Fe is within $\approx 10$\% of that for the larger basis.   
Thus, the $\gamma$SF depends mainly on the wavefunction properties and not strongly on the level density.
For the following discussion, we restrict 
ourselves to the excitation-energy range $ 5.8 \leq E_i \leq 8.0 $ MeV for $^{56}$Fe, and $ 5.0 \leq E_i \leq 8.0 $ MeV for $^{57}$Fe.

\begin{figure}[tb]
 \begin{center}
 \includegraphics[clip,width=0.95\columnwidth]{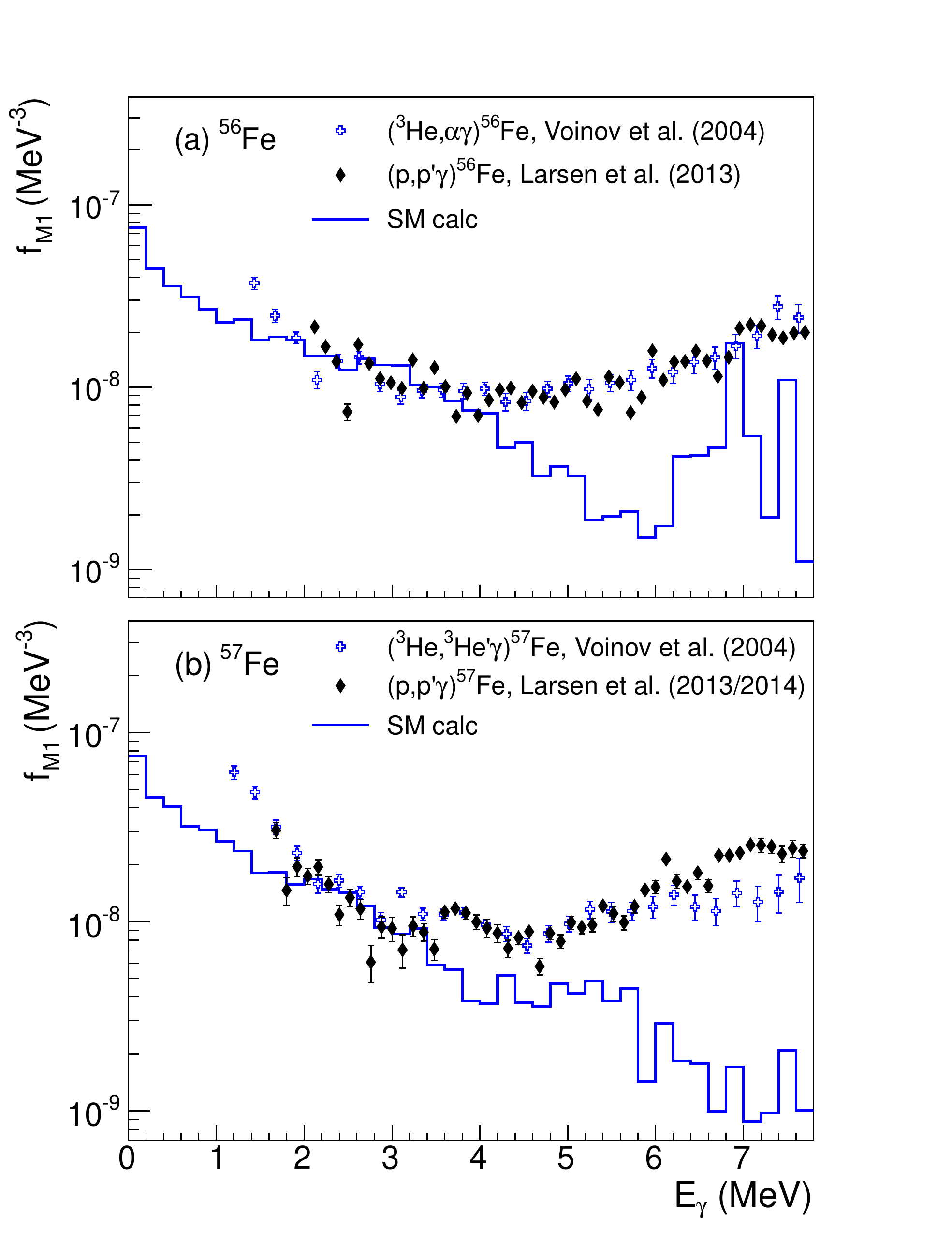}
 \caption {(Color online) Shell-model $M1$ strength functions (thick, blue lines) 
 for (a) $^{56}$Fe and (b) $^{57}$Fe are compared to 
 experimental data from Refs.~\cite{voinov2004,larsen2013}.  } 
 \label{fig:gsf}
 \end{center}
 \end{figure}

Following the analysis of Ref.~\cite{schwengner2013}, the $M1$ $\gamma$SF, $f_{M1}(E_\gamma)$, 
is obtained from the average $B(M1)$ values by
\begin{equation}
f_{\mathrm{M1}}(E_\gamma) = \frac{16\pi}{9(\hbar c)^3} \left<{B}(M1)\right> \rho_i(M1),
\end{equation} 
where the constant  $16\pi/9(\hbar c)^3 = 11.5473\times 10^{-9}$ $\mu_{N}^{-2}$MeV$^{-2}$, $\left<{B}(M1)\right>$ is given 
in units of $\mu_N^2$ and $\rho_i$ is the density of levels (in MeV$^{-1}$) having at least one $M1$ transition  
at the initial excitation energy $E_i$.  The resulting shell-model $M1$ strength functions are compared to data in Fig.~\ref{fig:gsf}. 
Clearly, the $f_{M1}$ component is strongly increasing as the $\gamma$-ray energy decreases, reproducing the trend observed 
in the data. Some discrepancy with the $^3$He-induced data~\cite{voinov2004} is observed for the low-energy transitions 
($E_\gamma\lesssim 1.8$ MeV). However, experimental and methodical difficulties prevented the extraction of data below 
$E_\gamma \approx 2$ MeV for the $(p,p')^{56,57}$Fe~\cite{larsen2013,larsen2014}. These problems would be expected also for 
the $^{3}$He-induced reactions, giving less confidence in these data points. 

 \begin{figure}[!htb]
 \begin{center}
 \includegraphics[clip,width=0.95\columnwidth]{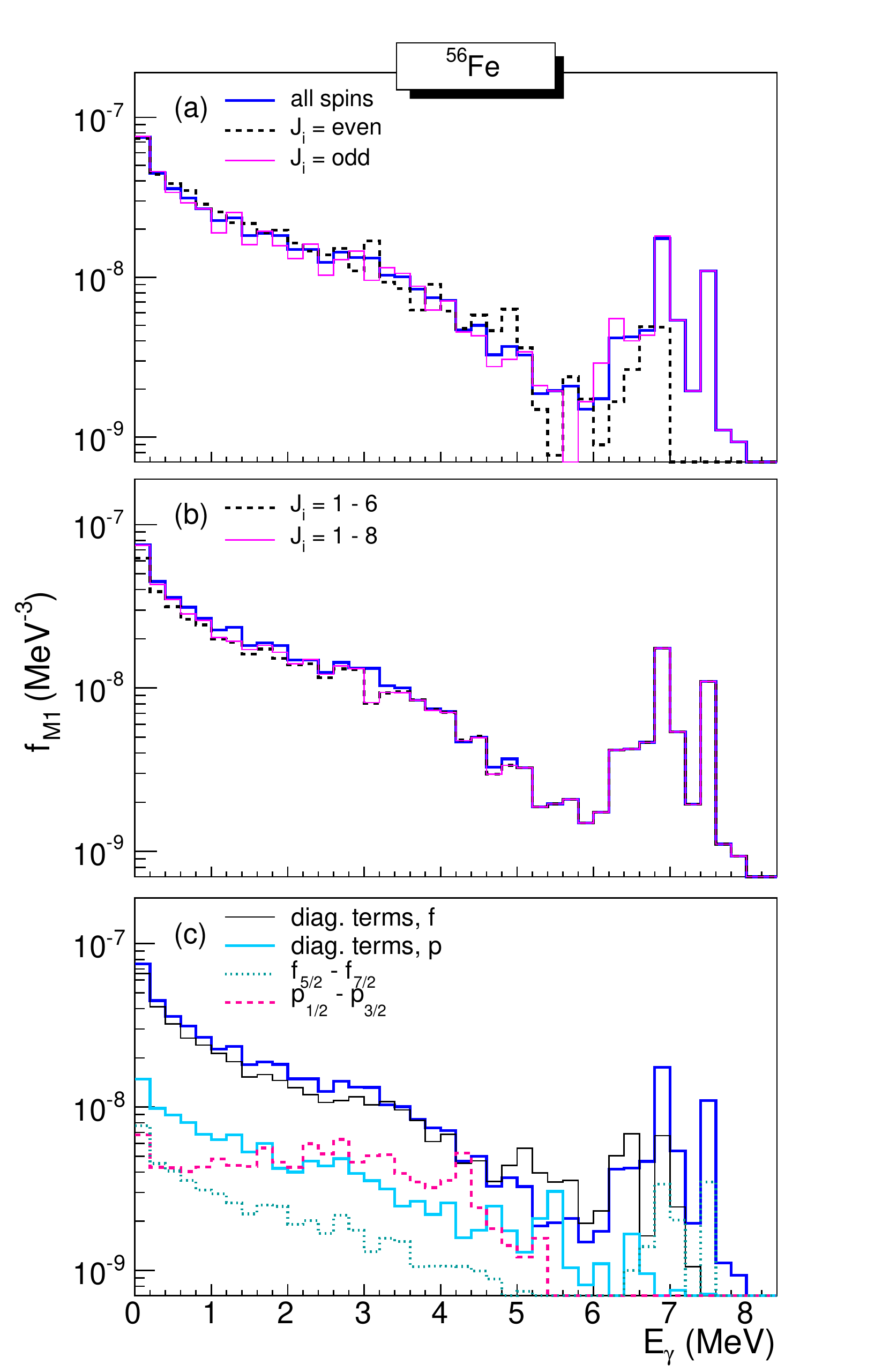}
 \caption {(Color online) Shell-model $M1$ strength functions of $^{56}$Fe for all spins and the 
 full $pf$ (thick, cyan line) 
 compared to the $\gamma$SF from (a) only even (dashed line) or only odd (magenta line) initial spins, 
 (b) initial spins $J_i = 1-6$ (dashed line) and $1-8$ (magenta line), and (c) $f_{M1}$ contributions for the 
 $  f_{7/2}-f_{5/2}  $ (dotted, cyan line),
 $  p_{3/2}-p_{1/2}  $ (dashed, pink line), and the  $  p_{1/2}-p_{1/2}$ and $ p_{3/2}-p_{3/2}  $ (light blue line),
 and the  $  f_{7/2}-f_{7/2}$ and $ f_{5/2}-f_{5/2}  $ (black line).  } 
 \label{fig:spin}
 \end{center}
 \end{figure}

There is a very interesting question whether this low-energy enhancement is related to the populated spin range of the initial
excited levels. Experiments on $^{95}$Mo using the two different charged-particle reactions 
($^{3}$He,$\alpha\gamma$)$^{95}$Mo~\cite{guttormsen2005}
and ($d,p$)$^{95}$Mo~\cite{wiedeking2012}, led to very similar shapes of the $\gamma$SF, although the $^3$He-induced pick-up reaction
is expected to populate higher spins on average, due to its preference for high-$\ell$ transfer (see, e.g., Ref.~\cite{casten1972}).
Furthermore, the shell-model calculations of Ref.~\cite{schwengner2013} gave that on average, all the included initial spins ($J_{i} = 0-6$) 
contributed approximately the same to the low-energy enhancement. 
The Brink hypothesis~\cite{brink1955} implies that
the strength function is independent of spin and excitation energy and is the same as that for the ground state.
In our calculations the $M1$ strength function for the ground state ($E_i=E_\gamma$ in Fig.~\ref{fig:BM1_matrices})
is dominated by a spin-flip resonance around 7 MeV.
When considering decay to all available levels at high excitation energies, the $M1$ strength function is broad  
with a significant low-energy component. This result was
also found for the Gamow-Teller strength function at high excitation in the $sd$ shell basis ~\cite{misch}.
Thus, we should think of a ``modified" Brink hypothesis where the strength is different from that
of the ground state, and becomes independent of excitation energy and spin above some given excitation energy.

To address the possible spin dependence, we have deduced the $\gamma$SF of $^{56}$Fe for various restrictions of the initial spin, 
see Fig.~\ref{fig:spin}. It is remarkable how persistent the low-energy enhancement is, regardless of the imposed spin restrictions.  
Specifically, we find no large deviations whether the initial spins are even or odd, or for the spin ranges 
resembling the experimental situation for ($p,p'$)$^{56}$Fe ($J_i \approx 1-6$~\cite{larsen2013}) or 
($^3$He,$\alpha$)$^{56}$Fe ($J_i \approx 1-8$~\cite{voinov2004}). 
Also tested (but not shown) are the strong restrictions $J_i=0-4$ and $J_i=5-10$. Again, the low-energy part ($E_\gamma < 3$ MeV)
remains largely unaffected. 
We therefore conclude that the $M1$ $\gamma$SF is, at least in this case, 
not very sensitive to the initial spin distribution, in agreement with experiments and the modified Brink hypothesis. 

Another and equally intriguing aspect of the Brink hypothesis is that the $\gamma$SF is assumed to be independent on
excitation energy. Experimentally, this has been investigated by extracting the $\gamma$SF from different excitation-energy
ranges (e.g. in Ref.~\cite{voinov2004}). We do the same test here by deducing the average $\gamma$SF for two different 
excitation-energy regions.
We find that the shapes of the $\gamma$SF's agree surprisingly well, and even the absolute strength is typically within
10\% for each 200-keV bin for $E_\gamma<3$ MeV. This fact further supports the experimental findings, and is again 
corroborating a modified Brink hypothesis.  

To understand the origin of the low-energy $M1$ strength, we restricted
the $M1$ matrix elements to the simple orbital combinations as shown in
Fig.~\ref{fig:spin}c.
From this we find that the high $\ell=f$ 
diagonal terms are most important for the lowest energy $M1$. 
It is very likely that these types of ``0$\hbar\omega$" diagonal terms 
with high $\ell$ should contribute to $M1$ spectra in all nuclei. 
This mechanism is different than the suggestion made in Ref.~\cite{schwengner2013} 
that the $M1$ low energy enhancement 
has a similar origin as the shears mode. In this mass region the shears mode comes from 
configurations with $  f_{7/2} $ proton holes and $  g_{9/2} $ neutron particles.
These are not in our model space, but should not be important until higher excitation
energies. 
 For $E1$, in contrast, there are no diagonal terms due to the
parity change, and the strong matrix elements involve those
in the giant-dipole resonance with a transition energy on the
order of 1$\hbar\omega$. In deformed nuclei the  ``0$\hbar\omega$" diagonal terms are responsible
for the orbital $M1$ "scissors"  mode~\cite{heyde2010} observed experimentally 
in the low-energy $\gamma$SF of deformed
heavy nuclei~\cite{guttormsen2014}.

\begin{figure}[tb]
 \begin{center}
 \includegraphics[clip,width=0.9\columnwidth]{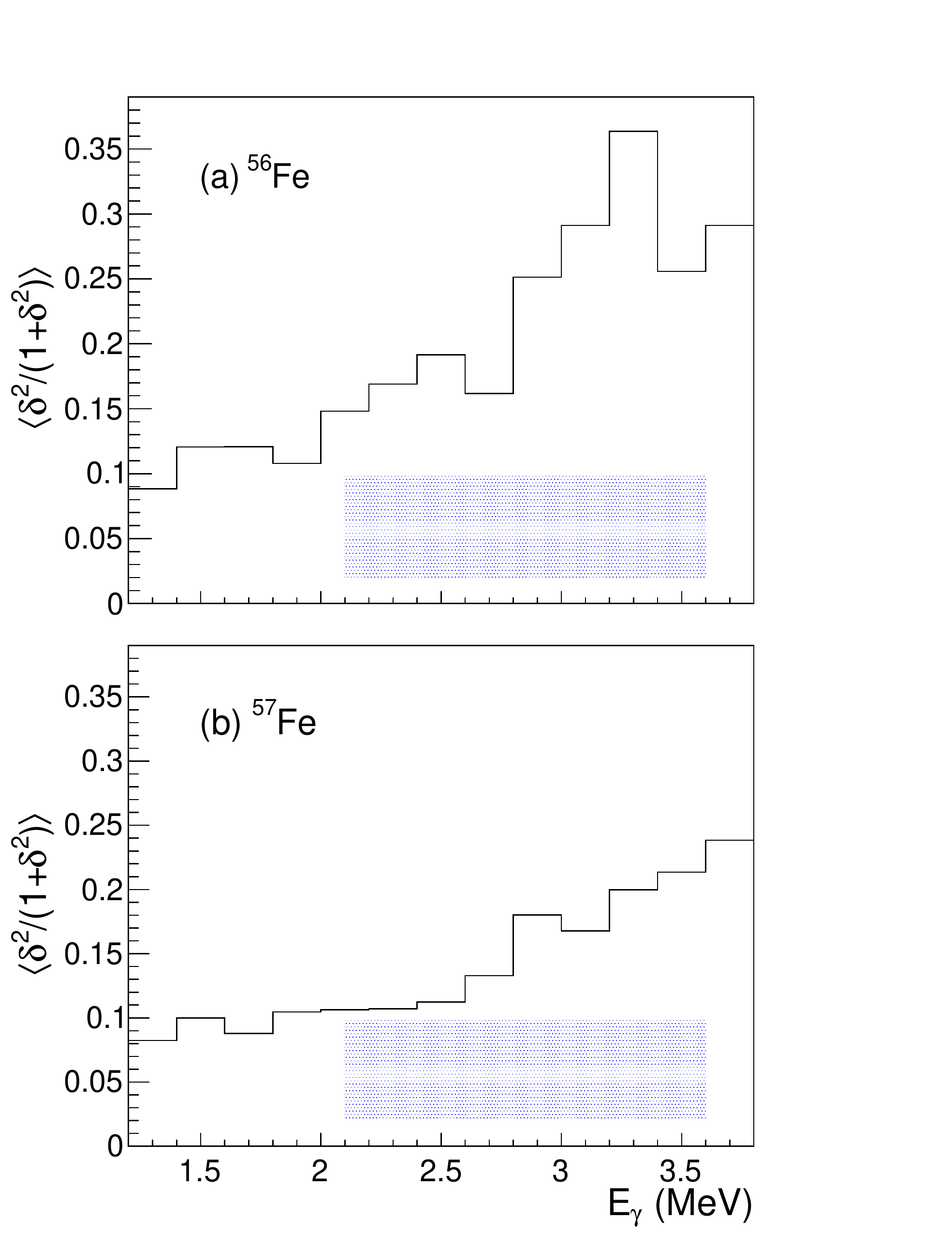}
 \caption {(Color online) Distribution of the $E2$ fraction for (a)  $^{56}$Fe and (b) $^{57}$Fe. The shaded area indicates 
 the experimentally investigated region for $^{56}$Fe in Ref.~\cite{larsen2013}.} 
 \label{fig:delta}
 \end{center}
 \end{figure}

To investigate the impact of the $E2$ transitions, we consider the average fraction of the $E2$'s to the transition rates
given by $\delta^2/(1+\delta^2)$, for which we obtain distributions as shown in Fig.~\ref{fig:delta}. 
We see that indeed, there is a certain contribution from $E2$ transitions, and in fact mostly so for $\gamma$-ray energies larger 
than the experimental low-energy enhancement region. From our calculations, we find that the average contribution from $E2$'s is $\approx 22$\% 
and $\approx 15$\% for $^{56,57}$Fe, respectively 
which is somewhat more than
the experimental findings of about $10$\% from angular distributions~\cite{larsen2013}. However, a possible low-energy $E1$ contribution in 
the experimental data might lead to a lower $E2$ fraction. 

Moreover, the influence of stretched ($\Delta J = 1$) versus non-stretched ($\Delta J = 0$) 
$M1$'s is studied by extracting the $\gamma$SF for these transitions separately. On average, the stretched transitions give 
$\approx 30$\% stronger $B(M1)$'s than the non-stretched in the case of $^{56}$Fe. This corresponds very well with experimental 
observations from angular-distribution measurements~\cite{larsen2013}. For $^{57}$Fe, the calculations 
indicate that the non-stretched transitions dominate by $\approx 20$\%.

In summary, we have performed large-basis shell-model calculations of $^{56,57}$Fe, which clearly give a large $M1$ strength for low 
$\gamma$-ray energies and at high excitation energies. 
The shell-model $f_{M1}$ functions are in excellent agreement with experimental data, and provide 
an explanation for the observed low-energy enhancement. Furthermore, restrictions on the $M1$ matrix elements
clearly show that 0$\hbar\omega$ transitions are responsible for the large low-lying strength. As this type of transitions
should be present for all nuclei, such a low-energy enhancement would be expected throughout the nuclear chart. Its presence
may significantly increase astrophysical ($n,\gamma$) reaction rates crucial for the understanding of the r-process.

\begin{acknowledgments}
We acknowledge support from NSF grant PHY-1068217 and PHY-1404442.
Computational work in support of this research was performed at Michigan State
University's High Performance Computing Facility.
A.~C.~L. acknowledges support from the Research Council of Norway, grant no. 205528. 
\end{acknowledgments}

\end{document}